# Extrinsic ferroelectricity originated from oxygen vacancy drift in HfO$_2$-based films

Yong Cheng,[1,2,§] Maoyuan Zheng,[1,3,§] Xingwang Zhang,[1,2,4] Hao Dong,[1,2] Yitian Jiang,[1,4] Jinliang Wu,[1] Jing Qi,[3] and Zhigang Yin[1,2,*]

[1]Key Lab of Semiconductor Materials Science, Institute of Semiconductors, Chinese Academy of Sciences, Beijing 100083, China.

[2]Center of Materials Science and Optoelectronics Engineering, University of Chinese Academy of Sciences, Beijing 100049, China.

[3]School of Materials and Energy, Lanzhou University, Lanzhou 730000, China.

[4]Joint Lab of Digital Optical Chip, Wuyi University, Jiangmen 529020, China.

---

[*]Corresponding author: yzhg@semi.ac.cn.

[§]These authors contributed equally to this work.

**Abstract:**

It is generally accepted that oxygen vacancies ($V_O$) play a central role in the emergence of ferroelectricity for $HfO_2$-based materials, but the underlying mechanism still remains elusive. Herein, starting from the basic characterization circuit, we propose that the observed ferroelectricity is extrinsic. A key finding is that charged $V_O$ oscillate within the sample under repeated electric pulses, yielding a nonlinear current which behaves similarly to the polarization current for a normal ferroelectric. This unwanted current signal results in a ferroelectric-like hysteresis loop with both remnant polarization and coercive field in good agreements with experimental values, given a charged $V_O$ concentration in the vicinity of $1\times10^{20}/cm^3$. Moreover, it is possible to exploit this mechanism to reproduce the effects of wake-up, split-up and limited endurance that are of crucial relevance for the device applications.

HfO$_2$-based ferroelectric films have attracted considerable research interest in nonvolatile memories and negative capacitance field-effect transistors, due to their scalability and compatibility with the complementary metal-oxide-semiconductor process [1-3]. However, the origin of the observed ferroelectricity is still under debate. In the majority of published works, the emergence of ferroelectricity has been attributed to the metastable orthorhombic (o) phases [4-6]. However, pure o-phase film cannot be available at present and its intrinsic properties are not well understood [7]. Moreover, huge polarization signals were also found in the weekly polar rhombohedral (r) epitaxial films [8,9], nonpolar cubic phase [10] and amorphous dielectric layers [11]. The reported ferroelectric properties strongly depend on the preparation route, and in most cases a post treatment in nitrogen atmosphere which favors for forming oxygen vacancies (V$_O$) is required [12]. Very recently, Nukala *et al.* nicely demonstrated that the ferroelectric switching is strongly intertwined with V$_O$ migration under electric field [13]. A variety of models such as flexoelectric effect accompanied with Vegard stress and V$_O$-assisted phase transition were proposed to uncover the role of V$_O$ in creating ferroelectricity [14,15], but no consensus has been achieved. Other key issues concerning HfO$_2$-based ferroelectrics are the effects of wake-up, split-up and poor endurance [16-18]. Although considerable efforts have been spent on these phenomena [16-20], the underlying mechanisms still remain unclear. The classification of the root cause of ferroelectricity is helpful for providing deeper insights into these effects.

Experimentally, the appearance of a polarization-electric field (***P-E***) hysteresis loop is generally viewed as a fingerprint of ferroelectrics and is used to identify ferroelectricity [21]. The characterization of ***P-E*** loop is commonly based on the Sawyer-Tower circuit (or its modified version), where the tested polarization signal is obtained by integrating the current (***I***) through a reference capacitor (or resistor) in series with the sample [21]. However, previous reports show that the occurrence of ***P-E*** loop is not necessarily associated with ferroelectricity [22]. For example, hysteresis loops were predicted for nonferroelectric systems containing metal-semiconductor Schottky contacts with heavily defective interfacial regions [23]. As to the unusual ferroelectricity in $HfO_2$-based materials, the door is kept open for an extrinsic origin.

Here we focus on, from the measurement point of view, the essential connection between $V_O$ and the ferroelectric-like behavior. We show that under repeated triangular pulses, the reversible drift of charged $V_O$ itself can generate a ***P-E*** hysteresis loop. This is because the charged $V_O$ transport is balanced by an electron flow in the external circuit, which behaves similarly with the polarization current of a normal ferroelectric. When the charged $V_O$ concentration is on the order of $10^{19}$-$10^{20}$/cm$^3$, the derived ***P-E*** loop is in line with experimental observations. The switchable polarization found in $HfO_2$-based materials is more likely an experimental artifact rather than an intrinsic property.

The $HfO_2$ sample containing charged $V_O$ can be regarded as a lossy capacitor in the ***P-E*** curve measurement. Its equivalent circuit is shown in Fig. 1(a), in which an ideal dielectric capacitor is shunted with an ionic resistor. Under the triangular pulse of electric field presented in Fig. 1(b), the measured current in the external circuit has two components: the dielectric displacement current ($I_\varepsilon$) and the current induced by ion drift between the top and bottom electrodes ($I_d$). To numerically derive the *P-E* loop based on the circuit shown in Fig. 1(a), we assume: i) The lateral dimension of the electrode/HfO$_2$/electrode sandwich structure is far larger than the HfO$_2$ film thickness so that the drift filed is uniform; ii) no interaction exists between neighboring charged V$_O$, and the annihilation, pinning and clustering of V$_O$ are not included; and iii) charged V$_O$ are homogeneously distributed within the film before the electric pulse is applied.

According to the Shockley-Ramo theorem [24,25], the drifting ion induced $I_d$ is written as

$$\boldsymbol{I}_d(t) = \sum_{i=1}^{n} Q \boldsymbol{v}_i(t) E_w \tag{1}$$

where *n* is the total amount of ions, *Q* is the ion charge, and $\boldsymbol{v}_i$ and $E_w$ are the instantaneous drift velocity of the *i*-th ion and the weighting field, respectively. For a two-electrode parallel plate, the weighting field is defined as $E_w = 1/d$ [24,25]. For the single-barrier drift, the drift velocity of the *i*-th ion under the drift field $\boldsymbol{E}(t)$ is [26]

$$\boldsymbol{v}_i(t) = \begin{cases} fae^{-\frac{E_a}{k_B T}} \sinh\left(\frac{Qa\boldsymbol{E}(t)}{2k_B T}\right) & \left(\frac{mt_0}{4} \leq t \leq \frac{mt_0}{4} + t_i, m \in N\right) \\ 0 & \left(\frac{mt_0}{4} + t_i < t < \frac{(m+1)t_0}{4}, m \in N\right) \end{cases} \tag{2}$$

where *f* is the jump frequency, *a* is the hopping distance, $E_a$ is the activation barrier, $k_B$ is the Boltzmann constant, *T* is the temperature, and $t_i$ is the drifting time of the *i*-th ion in a quadrant. The detected polarization signal is expressed as

$$\boldsymbol{P}(t) = \boldsymbol{P}_\varepsilon(t) + \boldsymbol{P}_d(t) = \frac{1}{A}\int_0^t [\boldsymbol{I}_\varepsilon(t) + \boldsymbol{I}_d(t)]dt \tag{3}$$

where *A* is the electrode area, and $\boldsymbol{I}_\varepsilon(t)$ corresponds to the time derivative of the displacement vector $\boldsymbol{D}(t)$, namely

$$\boldsymbol{I}_\varepsilon(t) = \frac{\partial}{\partial t}\boldsymbol{D}(t) = \varepsilon \frac{\partial}{\partial t}\boldsymbol{E}(t) \tag{4}$$

where $\varepsilon$ is the dielectric permittivity. The collective oscillation of ions under the triangular electric pulse is schematically illustrated in Fig. 1(c), and the corresponding time-dependent polarization signal and ***P*-*E*** curve can be derived on the basis of Eqs. (1)-(4).

We use r-phase $HfO_2$ as a model system. The activation energies for $V_O$ hoppings between all possible neighboring sites were calculated by density functional theory using the VASP code, and the details are provided in the Supplementary Material. An equivalent hexagonal cell is used [Fig. 2(a)], with its [001] direction corresponds to the r-$HfO_2$[111] orientation through index transformation. According to previous report [27], oxygen vacancies in r-$HfO_2$ are negative-U defects, i.e., the formation of singly charged $V_O$ ($V_O^{\bullet}$) is not energetically allowed. Moreover, our calculations show that for the same hopping, the activation barrier of neutral $V_O$ ($V_O^x$) is much larger than that of doubly charged $V_O$ ($V_O^{\bullet\bullet}$), implying $V_O^x$ can be seen as fixed defects (see the Supplementary Material). Therefore, only the drift of $V_O^{\bullet\bullet}$ is considered here. There exist four inequivalent O-atoms (categorized as III-1, III-2, IV-1 and IV-2), among which the former two are threefold coordinated, and the latter two are fourfold coordinated. The dashed arrows in Fig. 2(a) indicate the lowest-energy path for long-range migration, namely, $A_{III-1} \rightarrow B_{IV-1} \rightarrow C_{III-1}$ [A, B and C are the atom positions as illustrated in Fig. 2(a)], and the corresponding energy profile is shown in Fig. 2(b). The contributions from other migration paths are negligible, see the Supplementary Material.

The pathway shown in Fig. 2(b) has two barriers: one is 0.54 eV ($A_{III-1} \rightarrow B_{IV-1}$), and the other is 0.39 eV ($B_{IV-1} \rightarrow C_{III-1}$). According to Eq. (2), the latter one produces a drift velocity two orders of magnitude larger than the former. That is, the total drifting time across the $A_{III-1}\rightarrow B_{IV-1}\rightarrow C_{III-1}$ transition is approximately the time it takes to travel through $A_{III-1}\rightarrow B_{IV-1}$. Therefore, it is reasonable to simplify our simulation by assuming the $B_{IV-1}\rightarrow C_{III-1}$ transition as barrier-free [Fig. 2(b)]. When the electric field is applied along r-HfO$_2$[111] (the film normal), the average drift velocity for the two-barrier hopping can be modified as

$$v_i(t) = \begin{cases} fa_z^t e^{-\frac{E_a}{k_BT}} \sinh\left(\frac{Qa_zE(t)}{2k_BT}\right) & \left(\frac{mt_0}{4} \leq t \leq \frac{mt_0}{4} + t_i, m \in N\right) \\ 0 & \left(\frac{mt_0}{4} + t_i < t < \frac{(m+1)t_0}{4}, m \in N\right) \end{cases} \quad (5)$$

where $a_z$ and $a_z^t$ are the projections of the main $A_{III-1}\rightarrow B_{IV-1}$ hopping distance and the total $A_{III-1}\rightarrow B_{IV-1}\rightarrow C_{III-1}$ transition distance, onto the film normal, respectively.

The parameters used for our simulation are: $f = 1\times10^{14}$ Hz [28], $a_z^t = 2.95$ Å, $T = 298$ K, $Q = 2q$ ($q$ is the electron charge), $E_a = 0.54$ eV, $a_z = 1.49$ Å, $d = 10$ nm, $A = 1$ cm$^2$, $n = AdN = 1\times10^{14}$ ($N = 1\times10^{20}$/cm$^3$), $\varepsilon = 41\varepsilon_0 = 3.63\times10^{-10}$ F/m [29], $t_0 = 1$ ms, and $E_0 = 5$ MV/cm. Fig. 3(a) shows the derived time dependence of $I_d$ and $P_d$. In the first quadrant, all the mobile $V_O^{\bullet\bullet}$ transport upwardly until they arrive at the top of HfO$_2$ layer. Since $V_O^{\bullet\bullet}$ is assumed to be uniformly distributed within the film at $t = 0$, it can be obtained that the average drift distance of $V_O^{\bullet\bullet}$ in the first quadrant is $d/2$ [Fig. 1(c)]. Combining Eqs. (1) and (3), it immediately yields a saturation polarization of $qn/A$. Given a mobile $V_O^{\bullet\bullet}$ concentration of $1\times10^{20}$/cm$^3$, the derived saturation $P_d$ is 16 μC/cm$^2$, close to experimental values [12].

In the second quadrant, $V_O^{\bullet\bullet}$ stays at $x = d$ and therefore $I_d$ remains zero [Fig. 3(a)]. As the electric field is reversed (the third quadrant), $V_O^{\bullet\bullet}$ is driven away from the top, and $I_d$ has a minus sign and it peaks when $V_O^{\bullet\bullet}$ reaches the bottom. Similar to the second quadrant, the fourth quadrant is associated with a zero ion-drift current. In the fifth quadrant, $V_O^{\bullet\bullet}$ revert to the top and $I_d$ gradually increases until $V_O^{\bullet\bullet}$ arrive at the top. The $I_d$-$t$ curve is very similar to the time-dependent polarization current for a normal ferroelectric [21], and the resultant $P_d$-$t$ curve is also shown in Fig. 3(a). Since the electric field is linearly dependent on $t$ under the triangular waveform, we can easily transform the $I_d$-$t$ and $P_d$-$t$ data into $I_d$-$E$ and $P_d$-$E$ curves, as presented in Fig. 3(b). A distinct hysteresis is clearly visible in the $P_d$-$E$ curve, and the obtained coercive field ($E_c$) is ~3.3 MV/cm, in good agreement with the reported value for the 9-nm-thick r-phase film measured under a frequency of 1 kHz [8]. On the other hand, $I_\varepsilon$ has a constant magnitude and switches its sign when the sweep direction changes, and therefore $P_\varepsilon$ is linearly dependent on $t$ [Fig. 3(c)]. The superposition of the field-dependent $I_d$ and $I_\varepsilon$ signals corresponds to adding two peaks on the $I_\varepsilon$-$E$ curve, and the resultant $P$-$E$ loop is shown in Fig. 3(d). Since the obtained $P_\varepsilon$ value (18 μC/cm$^2$) at 5 MV/cm is comparable with the saturation $P_d$ (16 μC/cm$^2$), a slanted $P$-$E$ hysteresis is obtained. Such a hysteresis shape resembles those reported in most of the previously published works in the area of HfO$_2$-related ferroelectrics [12]. Using the positive-up-negative-down (PUND) method, the $P_\varepsilon$ component can be subtracted from the total signal, and a pure $P_d$-$E$ loop is experimentally available [8].

We show above that the ion charge transport induces a flow of electrons through the external circuit, which gives rise to a polarization signal. We emphasize that this polarization signal has nothing to do with electric dipoles, but is an experimental artifact.

The main features of the ion-drift-induced current are: i) it increases nonlinearly with the drift field; and ii) it vanishes when the ions arrive at the top (or bottom), until the applied bias is inverted. Due to this unusual behavior, the ion-drift current is indistinguishable from the polarization current of a ferroelectric − under repeated electric pulses, integrating either of them over time yields a hysteresis loop. Taking into account the oxygen-deficient nature of the HfO$_2$-based films [13], the appearance of a ***P-E*** hysteresis loop based on the Sawyer-Tower circuit or its advanced version, is not sufficient to demonstrate an intrinsic ferroelectricity. We note that the electrically driven oxygen vacancy drift in HfO$_2$ has indeed been observed by a variety of groups, through macroscopic [28,30] and microscopic [13] tools. Moreover, supposing a constant drift distance, the ion-drift-induced artificial polarization is proportional to the ion concentration. In previous reports, the remnant polarization ($P_r$) values are highly scattered which fall in between 5 and 50 μC/cm$^2$, corresponding to a mobile $V_O^{\bullet\bullet}$ concentration ranging from ~3×10$^{19}$/cm$^3$ to ~3×10$^{20}$/cm$^3$ in the case of a film thickness of 10 nm. Such a concentration is experimentally accessible for HfO$_2$ related materials, since similar or even higher values (up to 10$^{21}$/cm$^3$) have been reported previously [31-33].

In light of the above mechanism, we now discuss some details of the reported observations in HfO$_2$-based films. The first is the wake-up effect, which is featured by increased $P_r$ and $E_c$ upon field cycling. Although it was previously attributed to phase exchange [34] or domain de-pining [16] through V$_O$ redistribution, no consensus has been achieved as yet. On the basis of the drifting $V_O^{\bullet\bullet}$ model, this effect can be understood without involving the domain switching process. As shown in Fig. 4(a), in the $P_d$-$E$ loops the saturation value is proportional to the $V_O^{\bullet\bullet}$ concentration, while $E_c$ stays constant. Strikingly, the evolution of the $P$-$E$ loops by superposing these $P_d$-$E$ curves and the linear $P_\varepsilon$-$E$ signal [Fig. 4(b)], can perfectly reproduce the wake-up effect [8,10]. Upon increasing the $V_O^{\bullet\bullet}$ concentration from $1\times10^{19}/cm^3$ to $1\times10^{20}/cm^3$, the $P$-$E$ loop is gradually opened up with enhanced $P_r$ and $E_c$. In view of this, the wake-up effect can be interpreted simply by invoking the generation of $V_O^{\bullet\bullet}$ during electrical cycling. In fact, the electrical creation of oxygen vacancies has been reported in a variety of systems including $HfO_2$ [35-37]. According to first principles calculations, Strand *et al.* also demonstrated that strong electric field facilitates reducing the formation energy of $V_O$ in $HfO_2$ [38]. We strengthen that wake-up-free pristine samples can only be realized under extreme preparation conditions beneficial for generating $V_O$, such as high-temperature nitrogen annealing [39], low oxygen pressure sputtering [40], and epitaxial growth under ultra-high temperatures [41]. Also notable is that a $P$-$E$ loop evolution similar to Fig. 4(b) was indeed observed by Ryu *et al.*, through tuning the oxygen pressure during preparation [40].

The split-up phenomenon and the poor endurance of $HfO_2$-based films are also understandable based on our model. The split-up effect, characterized by the splitting of the current peaks in both the positive and negative bias regions in the $I$-$E$ loop [17], may be closely allied to the occurrence of multiple $V_O^{\bullet\bullet}$ drift channels (see the Supplementary Material). The coexistence of diverse phases or competing migration pathways within a single phase, allow the emergence of multiple current peaks and the resultant spilt-up. Further field cycling induces phase exchange [42] or redistribution of $V_O^{\bullet\bullet}$ [43] towards the most favorable channel, which may be relevant to the subsequent merging of these peaks [20]. On the other hand, an electric field of 3 MV/cm or above is needed to achieve a saturated *P-E* curve for $HfO_2$-based materials, leading to a considerable electronic leakage current after limited switching cycles [44]. Oxygen vacancies are electron traps [45], and the electron charge transport in $HfO_2$ is described by the phonon-assisted tunneling between adjacent traps [46]. The trapping of injected electrons converts part of $V_O^{\bullet\bullet}$ into $V_O^{\bullet}$ or $V_O^x$ [47] and, as a result, the collected $P_r$ signal is decreased which behaves similarly to the ferroelectric fatigue.

In conclusion, we propose it is the charged $V_O$ drift that causes the ferroelectric-like behavior for $HfO_2$-based materials. This mechanism is not specific to r-$HfO_2$, but can be transposed to other phases, of both the polycrystalline and single-crystalline film forms. The scenario we presented here naturally manifests itself in the experimental findings, such as the oxygen deficiency of $HfO_2$-based samples and the direct observations of $V_O$ migration. Based on our model, phenomena including scattered $P_r$, wake-up, split-up and poor endurance can also be understood in a simple and rational way.

This work was financially supported by the National Key Research and Development Program of China (2017YFB0405600), and the National Natural Science Foundation of China (62074145 and 61874112).

**REFERENCES**

[1] T. S. Boscke, J. Muller, D. Brauhaus, U. Schroder, and U. Bottger, Appl. Phys. Lett. 99, 102903 (2011).

[2] X. Xu, F.-T. Huang, Y. Qi, S. Singh, K. M. Rabe, D. Obeysekera, J. Yang, M.-W. Chu, and S.-W. Cheong, Nat. Mater. 20, 826 (2021).

[3] L. Q. Tu, R. R. Cao, X. D. Wang, Y. Chen, S. Q. Wu, F. Wang, Z. Wang, H. Shen, T. Lin, P. Zhou, X. J. Meng, W. D. Hu, Q. Liu, J. L. Wang, M. Liu, and J. H. Chu, Nat. Commun. 11, 101 (2020).

[4] T. D. Huan, V. Sharma, G. A. Rossetti, and R. Ramprasad, Phys. Rev. B 90, 064111 (2014).

[5] X. Sang, E. D. Grimley, T. Schenk, U. Schroeder, and J. M. Lebeau, Appl. Phys. Lett. 106, 162905 (2015).

[6] Y. Qi, S. Singh, C. Lau, F.-T. Huang, X. Xu, F. J. Walker, C. H. Ahn, S.-W. Cheong, and K. M. Rabe, Phys. Rev. Lett. 125, 257603 (2020).

[7] Z. Zhang, S.-L. Hsu, V. A. Stoica, H. Paik, E. Parsonnet, A. Qualls, J. Wang, L. Xie, M. Kumari, S. Das, Z. Leng, M. McBriarty, R. Proksch, A. Gruverman, D. G. Schlom, L.-Q. Chen, S. Salahuddin, L. W. Martin, and R. Ramesh, Adv. Mater. 33, 2006089 (2021).

[8] Y. Wei, P. Nukala, M. Salverda, S. Matzen, H. J. Zhao, J. Momand, A. S. Everhardt, G. Agnus, G. R. Blake, P. Lecoeur, B. J. Kooi, J. Iniguez, B. Dkhil, and B. Noheda, Nat. Mater. 17, 1095 (2018).

[9] L. Begon-Lours, M. Mulder, P. Nukala, S. De Graaf, Y. A. Birkholzer, B. Kooi, B.

Noheda, G. Koster, and G. Rijnders, Phys. Rev. Materials 4, 043401 (2020).

[10] S. Starschich, D. Griesche, T. Schneller, R. Waser, and U. Bottger, Chemical solution deposition of ferroelectric yttrium-doped hafnium oxide films on platinum electrodes. Appl. Phys. Lett. 104, 202903 (2014).

[11] Z. Feng, Y. Peng, Y. Shen, Z. Li, H. Wang, X. Chen, Y. Wang, M. Jing, F. Lu, W. Wang, Y. Cheng, Y. Cui, A. Dingsun, G. Han, H. Liu, and H. Dong, Adv. Electron. Mater. 2100414 (2021).

[12] M. H. Park, Y. H. Lee, H. J. Kim, Y. J. Kim, T. Moon, K. D. Kim, J. Muller, A. Kersch, Uwe Schroeder, T. Mikolajick, and C. S. Hwang, Adv. Mater. 27, 1811 (2015).

[13] P. Nukala, M. Ahmadi, Y. Wei, S. De Graaf, E. Stylianidis, T. Chakrabortty, S. Matzen, H. W. Zandbergen, A. Bjorling, D. Mannix, D. Carbone, B. Kooi, and B. Noheda, Science 372, 630 (2021).

[14] M. D. Glinchuk, A. N. Morozovska, A. Lukowiak, W. Stręk, M. V. Silibin, D. V. Karpinsky, Y. Kim, S. V. Kalinin, J. Alloys Compd. 830, 153628 (2020).

[15] K. Z. Rushchanskii, S. Blugel, and M. Lezaic, Phys. Rev. Lett. 127, 087602 (2021).

[16] D. Zhou, J. Xu, Q. Li, Y. Guan, F. Gao, X. Dong, J. Muller, T. Schenk, and U. Schoder, Appl. Phys. Lett. 103, 192904 (2013).

[17] T. Schenk, U. Schroeder, M. Pesic, M. Popovici, Y. V. Pershin, and T. Mikolajick, ACS Appl. Mater. Interfaces 6, 19744 (2014).

[18] J. Lyu, I. Fina, and F. Sanchez, Appl. Phys. Lett. 117, 072901 (2020).

[19] S. S. Fields, S. W. Smith, P. J. Ryan, S. T. Jaszewski, I. A. Brummel, A. Salanova,

G. Esteves, S. L. Wolfley, M. D. Henry, P. S. Davids, and J. F. Ihlefeld, ACS Appl. Mater. Interfaces 12, 26577 (2020).

[20] T. Schenk, M. Hoffmann, J. Ocker, M. Pesic, T. Mikolajick, and U. Schroeder, ACS, Appl. Mater. Interfaces 7, 20224 (2015).

[21] T. Schenk, E. Yurchuk, S. Muller, U. Schroeder, S. Starschich, U. Bottger, and T. Mikolajick, Appl. Phys. Rev. 1, 041103 (2014).

[22] B. Kim, D. Seol, S. Lee, H. N. Lee, and Y. Kim, Appl. Phys. Lett. 109, 102901 (2016).

[23] L. Pintilie and M. Alexe, Appl. Phys. Lett. 87, 112903 (2005).

[24] S. Ramo, Proc. IRE 27, 584 (1939).

[25] K. A. Recine, J. B. R. Battat, and S. Henderson, Am. J. Phys. 82, 322 (2014).

[26] D. B. Strukov and R. S. Williams, Appl. Phys. A 94, 515 (2009).

[27] M. Y. Zheng, Z. G. Yin, Y. Cheng, X. W. Zhang, J. L. Wu, and J. Qi, Appl. Phys. Lett. 119, 172904 (2021).

[28] S. Zafar, H. Jagannathan, L. F. Edge, and D. Gupta, Appl. Phys. Lett. 98, 152903 (2011).

[29] The relative dielectric permittivity, 41, is extracted from the ***P-E*** loop for the r-phase $Hf_{0.5}Zr_{0.5}O_2$ film, see Ref. [8].

[30] T. Nagata, M. Haemori, Y. Yamashita, H. Yoshikawa, Y. Iwashita, K. Kobayashi, and Chikyow, Appl. Phys. Lett. 97, 082902 (2010).

[31] E. Hildebrandt, J. Kurian, and L. Alff, J. App. Phys. 112, 114112 (2012).

[32] D. R. Islamov, V. A. Gritsenko, C. H. Cheng, and A. Chin, Appl. Phys. Lett. 105,

222901 (2014).

[33] N. Capron, P. Broqvist, and A. Pasquarello, Appl. Phys. Lett. 91, 192905 (2007).

[34] S. S. Fields, S. W. Smith, P. J. Ryan, S. T. Jaszewski, I. A. Brummel, A. Salanova, G. Esteves, S. L. Wolfley, M. D. Henry, P. S. Davids, and J. F. Ihlefeld, ACS Appl. Mater. Interfaces 12, 26577 (2020).

[35] J. Jeong, N. Aetukuri, T. Graf, T. D. Schladt, M. G. Samant, and S. S. P. Parkin, Science 339, 1402 (2013).

[36] A. L. Mariano and R. Poloni, J. Chem. Phys. 154, 224703 (2021).

[37] C. Li, Y. Yao, X. Shen, Y. Wang, J. Li, C. Gu, R. Yu, Q. Liu, and M. Liu, Nano Res. 8, 3571 (2017).

[38] J. W. Strand, J. Cottom, L. Larcher, and A. L. Shluger, Phys. Rev. B 102, 014106 (2020).

[39] M. H. Park, H. J. Kim, Y. J. Kim, W. Lee, T. Moon, and C. S. Hwang, Appl. Phys. Lett. 102, 242905 (2013).

[40] T.-H. Ryu, D.-H. Min, and S.-M. Yoon, J. Appl. Phys. 128, 074102 (2020).

[41] J. Lyu, I. Fina, R. Solanas, J. Fontcuberta, and F. Sanchez, Appl. Phys. Lett. 113, 082902 (2018).

[42] S. S. Fields, S. W. Smith, P. J. Ryan, S. T. Jaszewski, I. A. Brummel, A. Salanova, G. Esteves, S. L. Wolfley, M. D. Henry, P. S. Davids, and J. F. Ihlefeld, ACS Appl. Mater. Interfaces 12, 26577 (2020).

[43] S. Starschich, S. Menzel, and U. Bottger, Appl. Phys. Lett. 108, 032903 (2016).

[44] M. H. Park, Y. H. Lee, T. Mikolajick, U. Schroeder, and C. S. Hwang, MRS

Commun. 8, 795 (2018).

[45] G. Bersuker, J. H. Sim, C. S. Park, C. D. Young, S. V. Nadkarni, R. Choi, and B. H. Lee, IEEE Trans. Device Mater. Rel. 7, 138 (2007).

[46] V. A. Gritsenko and A. A. Gismatulin, Appl. Phys. Lett. 117, 142901 (2020).

[47] R. A. Marcus, Rev. Mod. Phys. 65, 599 (1993).

**FIGURE CAPTIONS**

**FIG. 1.** (a) The simplified Sawyer-Tower circuit for *P-E* measurement; the dashed blue rectangle denotes the equivalent circuit for the HfO$_2$ film containing charged V$_O$, where a dielectric capacitor (*C*) is shunted with an ionic resistor ($R_i$); the reference capacitor ($C_{ref}$) can be replaced by a resistor. (b) Schematic of the triangular pulse with amplitude of $E_0$ and period of $t_0$; the dots marked by (1), (2), (3), (4), (5) and (6) denote the electric filed at $t$ = 0, $t_0/4$, $t_0/2$, $3t_0/4$, $t_0$, and $5t_0/4$. (c) Sketch of the electric-field-driven collective oscillation of charged V$_O$. For convenience, the top electrode is assumed to be grounded.

**FIG. 2.** (a) Crystal structure of r-HfO$_2$ with *R3m* space group. (b) The energy profile of the most favorable long-range $V_O^{\bullet\bullet}$ migration path marked by dashed arrows in (a); the solid line shows the single-barrier approximation of the two-barrier transition.

**FIG. 3.** The derived $I_d$-*t* and $P_d$-*t* (a), $I_d$-*E* and $P_d$-*E* (b), $I_\varepsilon$-*t* and $P_\varepsilon$-*t* (c), and *I-E* and *P-E* (d) curves for r-HfO$_2$ upon a 1 kHz triangular pulse; the electrode area *A* is assumed to be 1 cm$^2$.

**FIG. 4.** Simulated $P_d$-*E* and $P_\varepsilon$-*E* (a), and *P-E* (b) curves of the 10-nm-thick r-HfO$_2$ sample with various $V_O^{\bullet\bullet}$ concentration; the inset in (b) displays the variations of $P_r$ and $E_c$ with respect to the $V_O^{\bullet\bullet}$ concentration.

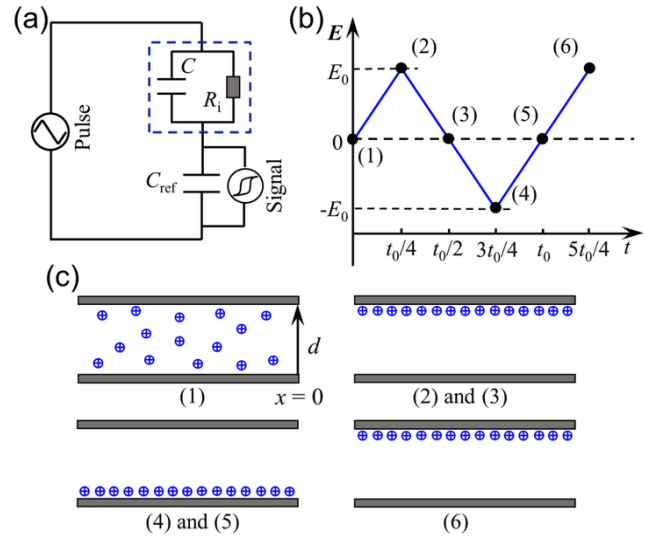

FIG. 1.

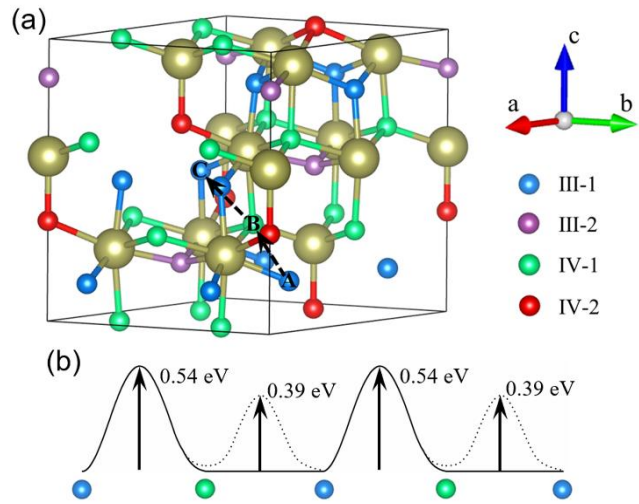

FIG. 2.

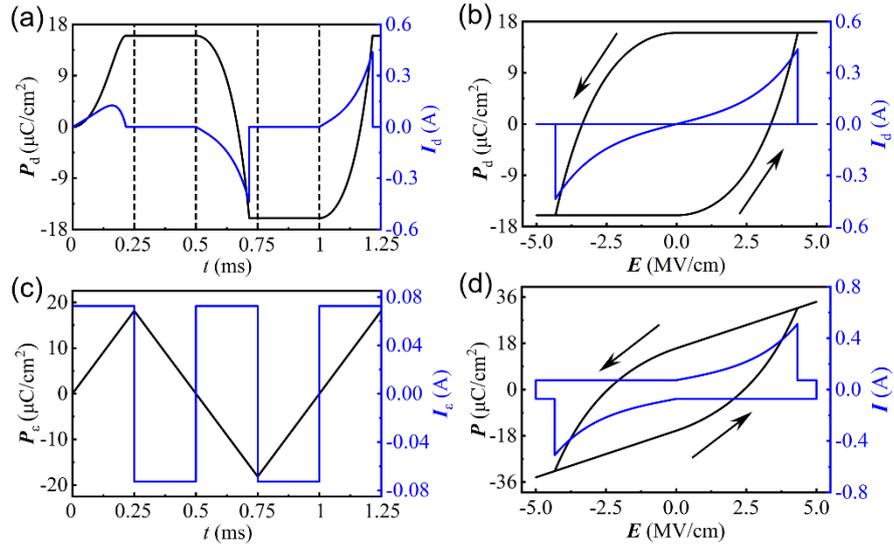

FIG. 3.

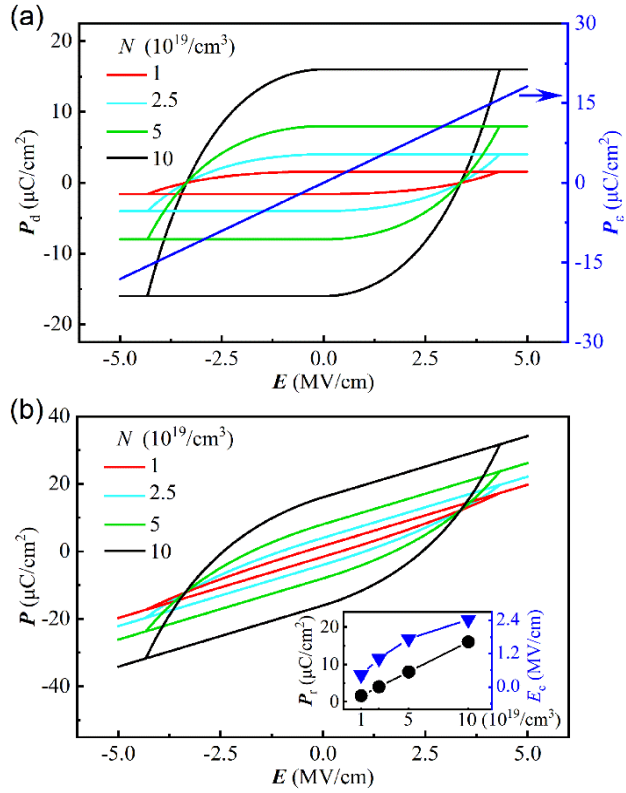

FIG. 4.

# Supplementary Material for

# Extrinsic ferroelectricity originated from oxygen vacancy drift in HfO$_2$-based films

Yong Cheng,[1,2,§] Maoyuan Zheng,[1,3,§] Xingwang Zhang,[1,2,4] Hao Dong,[1,2] Yitian Jiang,[1,4] Jinliang Wu,[1] Jing Qi,[3] and Zhigang Yin[1,2,*]

[1]Key Lab of Semiconductor Materials Science, Institute of Semiconductors, Chinese Academy of Sciences, Beijing 100083, China.

[2]Center of Materials Science and Optoelectronics Engineering, University of Chinese Academy of Sciences, Beijing 100049, China.

[3]School of Materials and Energy, Lanzhou University, Lanzhou 730000, China.

[4]Joint Lab of Digital Optical Chip, Wuyi University, Jiangmen 529020, China.

## List of Supplemental Material:

    I. Numerical details

    II. Activation barriers of $V_O^x$

    III. Polarization signal contributed by other drift paths

    IV. Simulation of the split-up effect

---

*Corresponding author: yzhg@semi.ac.cn.

§These authors contributed equally to this work.

# I. Numerical details

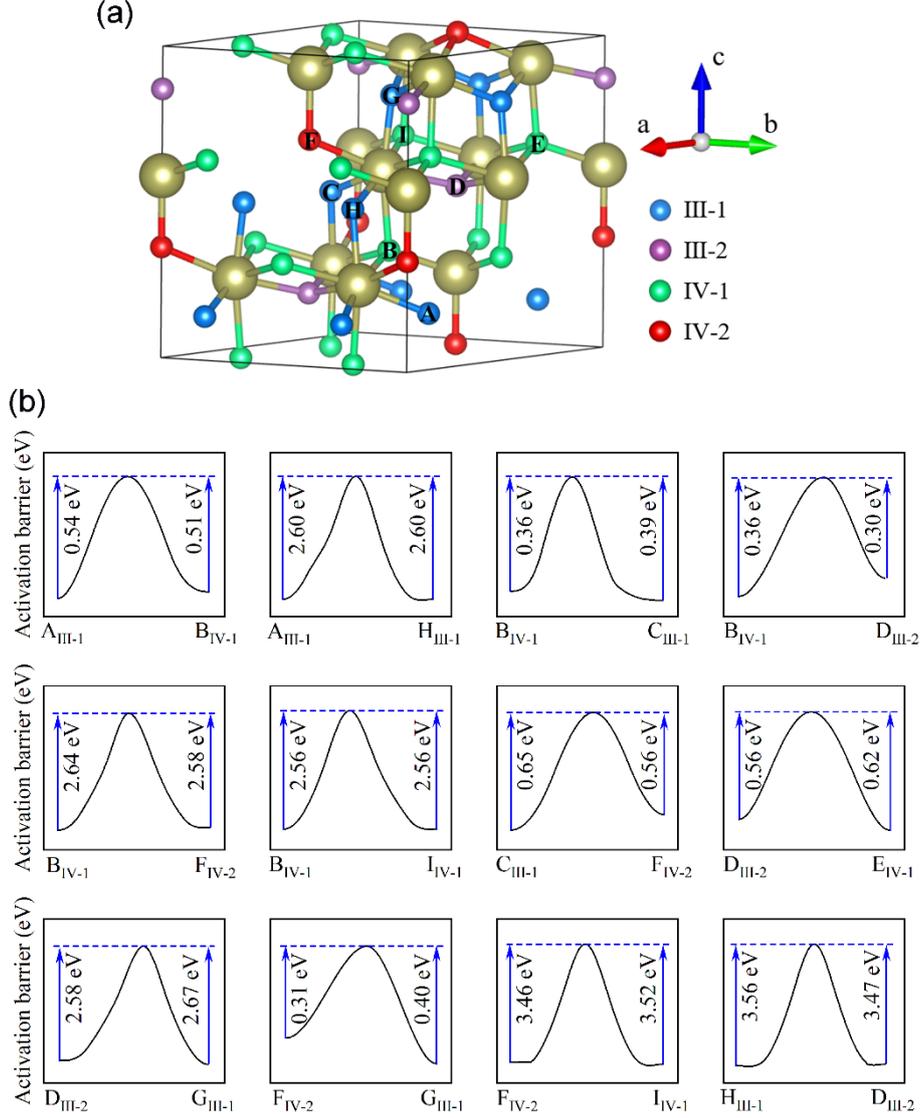

**FIG. S1** (a) Crystal structure of r-HfO$_2$ (space group: $R3m$); the four inequivalent O-atoms are labeled by blue, brown, green and red colors; A, B, …, and I denote the O-atom positions. (b) Activation energies for $V_O^{\bullet\bullet}$ hopping between neighboring sites; the most energetically favorable, long-range migration path is A$_{\text{III-1}}$→B$_{\text{IV-1}}$→C$_{\text{III-1}}$.

First principles calculations were performed by density functional theory (DFT) within the projector-augmented waves (PAW) [S1,S2] as implemented in the Vienna Ab initio Simulation Package (VASP) [S3]. The generalized gradient approximation (GGA) in the parametrization by Perdew-Burke-Ernzerhof (PBE) was used [S4]. We employ a cutoff energy of 550 eV for the plane wave expansion and a 5×5×4 $k$-point grid for Brillouin zone integrations. The structure of r-phase HfO$_2$ (space group: $R3m$) is optimized and the obtained lattice parameters are: $a = b = 7.21$ Å, $c = 8.84$ Å, $\alpha = \beta = 90°$ and $\gamma = 120°$ (in hexagonal coordinates), in good consistent with previous predictions by Zhang *et al* [S5].

The activation energies for $V_O^{\bullet\bullet}$ hoppings between all possible near-neighbor sites were calculated by the climbing image nudged elastic band (CI-NEB) method as implemented in VASP. In the CI-NEB calculations, a group of images are linearly interpolated between the initial and final states to represent the migration path [S6]. Three or five intermediate images are used in this study, and each image is relaxed until the maximum force per atom is less than 0.02 eV/Å. Since the electric field is assumed to be applied along the $c$ direction (the film normal) in Fig. S1(a), we do not consider the hopping transitions within the $a$-$b$ plane [the (111) plane in the rhombohedral coordinates]. The activation energies for all neighboring $V_O^{\bullet\bullet}$ hoppings are shown in Fig. S1(b), and the obtained lowest-energy pathway for long-range migration is $A_{III-1} \rightarrow B_{IV-1} \rightarrow C_{III-1}$.

## II. Activation barriers of $V_O^x$

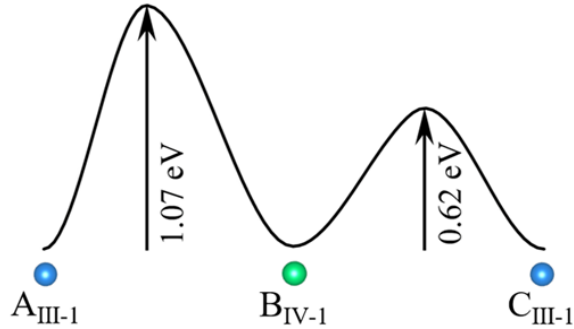

**FIG. S2** The energy profile for $V_O^x$ migration along the $A_{III-1} \rightarrow B_{IV-1} \rightarrow C_{III-1}$ path.

In this study, we also calculated the activation energies for neighboring transitions of $V_O^x$ in r-phase $HfO_2$. Fig. S2 shows the obtained energy profile along $A_{III-1} \rightarrow B_{IV-1} \rightarrow C_{III-1}$ (see Fig. S1). The $A_{III-1} \rightarrow B_{IV-1}$ and $B_{IV-1} \rightarrow C_{III-1}$ transitions are associated with activation barriers of 1.07 and 0.62 eV, respectively. By contrast, the $E_a$ values of $V_O^{\bullet\bullet}$ along the same path are only 0.54 and 0.39 eV. That is, for the same transition, the activation barrier of $V_O^x$ is about 2 times larger than that of $V_O^{\bullet\bullet}$ in r-$HfO_2$, akin to the observations in monoclinic $HfO_2$ [S7]. Combining the neutral nature and the rather high $E_a$ values, the drift of $V_O^x$ and its contribution to the polarization signal are not included in this work.

### III. Polarization signal contributed by other drift paths

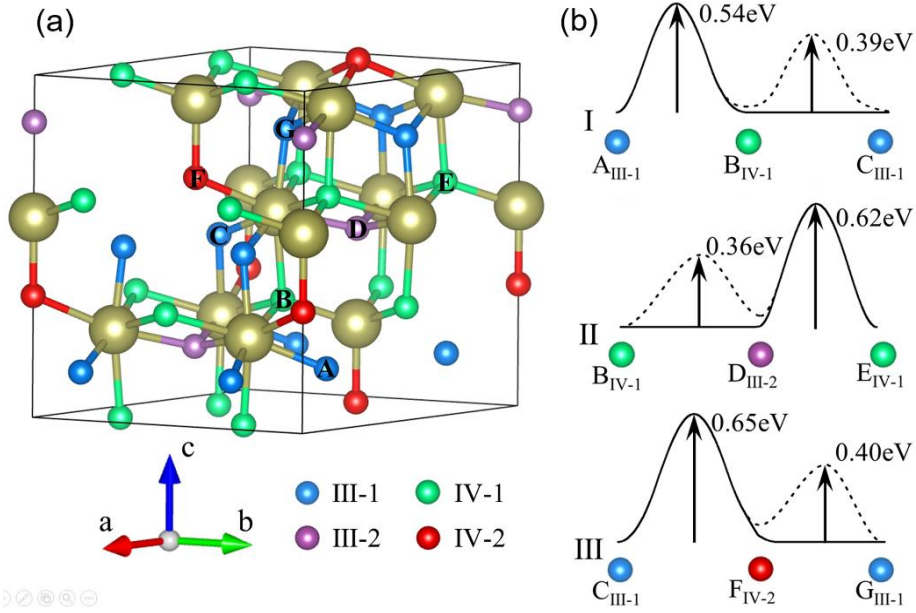

**FIG. S3** (a) Crystal structure of r-HfO$_2$ (space group: $R3m$). (b) The energy profiles of the three most favorable migration paths of $V_O^{\bullet\bullet}$, including I: A$_{III-1}$→B$_{IV-1}$→C$_{III-1}$, II: B$_{IV-1}$→D$_{III-2}$→E$_{IV-1}$, and III: C$_{III-1}$→F$_{III-2}$→G$_{III-1}$; the solid lines are the single-barrier approximations of these two-barrier transitions.

According to the results shown in Fig. S1(b), the three most favorable long-range migration paths are: A$_{III-1}$→B$_{IV-1}$→C$_{III-1}$ (path I), B$_{IV-1}$→D$_{III-2}$→E$_{IV-1}$ (path II), and C$_{III-1}$→F$_{III-2}$→G$_{III-1}$ (path III). Here A, B, …, and G are the O-atom positions, as shown in Fig. S3(a). The energy profiles of these paths are presented in Fig. S3(b). In r-HfO$_2$, the O-atoms on sites III-2 and IV-2 as a whole account for 25% of the total oxygen atoms. Assuming 25% of the total O-atoms are migrated along paths II and III, our calculations show that the resultant polarization is only ~1% of the total signal. This is because the barrier heights of paths II and III are so large that the drift distance of $V_O^{\bullet\bullet}$ along the film normal is very limited (~0.6 nm and ~0.4 nm for paths II and III, respectively), under a 1 kHz triangular pulse. Moreover, the formation energies of $V_O^{\bullet\bullet}$ on sites III-2 and IV-2 are much higher than those of III-1 and IV-1, suggesting the amount of $V_O^{\bullet\bullet}$ on sites III-2 and IV-2 is less than 25% of the total $V_O^{\bullet\bullet}$. Additionally, $V_O^{\bullet\bullet}$ on sites III-2 and IV-2 can be transferred to path I under electric field, through $D_{III-2} \rightarrow E_{IV-1}$ and $F_{III-2} \rightarrow G_{III-1}$. Therefore, it is safe to ignore the polarization contributions from path II and path III.

## IV. Simulation of the split-up effect

The parameters used for the two-channel simulation are: $f = 1 \times 10^{14}$ Hz, $a_1 = a_2 = 2.50$ Å, $T = 298$ K, $Q = 2q$ ($q$ is the electron charge), $E_{a1} = 0.54$ eV, $E_{a2} = 0.49$ eV, $d = 10$ nm, $A = 1$ cm$^2$, $n_1 = n_2 = 5 \times 10^{13}$, $\varepsilon = 30\varepsilon_0 = 2.66 \times 10^{-10}$ F/m, $t_0 = 1$ ms, and $E_0 = 5$ MV/cm. For simplicity, we presume that the two channels share the same hopping distance and equal amounts of $V_O^{\cdot\cdot}$. Based on Eqs. (1)-(4) in the main text, we can derive the *I-E* and *P-E* curves, as displayed in Fig. S4(b). As expected, the current peaks in both the positive and negative bias regions are splitted into two. Resultantly, the *P-E* loop has a pinched shape and the split-up phenomenon is reproduced.

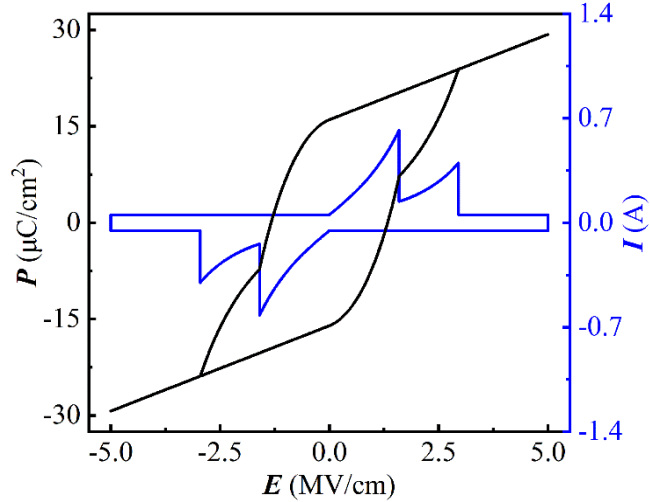

**FIG. S4** The simulated *I-E* and *P-E* curves for the two-channel system with $E_a$ values of 0.54 and 0.49 eV; the electrode area is assumed to be 1 cm$^2$.

The emergence of split-up process may correlate to the coexistence of diverse phases or competing migration pathways within a single phase. However, as shown in Sec. III, the long-range migration of $V_O^{\cdot\cdot}$ in the r-phase film is dominated by the lowest-energy path. Therefore, the split-up procedure was mainly found in the multiphase polycrystalline films [S8], but seldom reported for r-phase epitaxial layers [S9]. For the multiphase polycrystalline films, monoclinic (space group: $P2_1/c$), cubic (space group: $Fm3m$), tetragonal (space group: $P4_2/nmc$) and orthogonal (space group: $Pca2_1$) phases may coexist with each other. According to previous report, monoclinic HfO$_2$ has an $E_a$ of 0.49 eV for $V_O^{\bullet\bullet}$ along the lowest-energy, long-range migration path. We also calculated the activation energies of $V_O^{\bullet\bullet}$ for cubic, tetragonal and orthogonal HfO$_2$ along the most favorable pathways, and the obtained values are 0.30, 0.25 and 0.51 eV, respectively. Also notable is that the activation barrier for $V_O^{\bullet\bullet}$ hopping is extremely sensitive to the strain and dopant distribution [S10]. Therefore, in the multiphase polycrystalline films, it is not surprising to find the double or multiple splits of the current peaks.

**REFERENCES**

[S1]  P. E. Blochl, Phys Rev B 50, 17953 (1994).

[S2]  G. Kresse and D. Joubert, Phys Rev B 59, 1758 (1999).

[S3]  G. Kresse and J. Furthmüller, Phys Rev B 54, 11169 (1996).

[S4]  J. P. Perdew, K. Burke, and M. Ernzerhof, Phys. Rev. Lett. 77, 3865 (1996).

[S5]  Y. Zhang, Q. Yang, L. Tao, E. Y. Tsymbal, and V. Alexandrov, Phys. Rev. Appl. 14, 014068 (2020).

[S6]  G. Henkelman, B. P. Uberuaga, and H. Jonsson, J. Chem. Phys. 113, 9901 (2000).

[S7]  N. Capron, P. Broqvist, and A. Pasquarello, Appl. Phys. Lett. 91, 192905 (2007).

[S8]  S. S. Fields, S. W. Smith, P. J. Ryan, S. T. Jaszewski, I. A. Brummel, A. Salanova, G. Esteves, S. L. Wolfley, M. D. Henry, P. S. Davids, and J. F. Ihlefeld, ACS Appl. Mater. Interfaces 12, 26577 (2020).

[S9]  Y. Wei, P. Nukala, M. Salverda, S. Matzen, H. J. Zhao, J. Momand, A. S. Everhardt, G. Agnus, G. R. Blake, P. Lecoeur, B. J. Kooi, J. Iniguez, B. Dkhil, and B. Noheda, Nat. Mater. 17, 1095 (2018).

[S10] R. Pornprasertsuk, P. Ramanarayanan, C. B. Musgrave, and F. B. Prinz, J. Appl. Phys. 98, 103513 (2005).